\documentclass[smallextended]{svjour3}       
\smartqed  
\usepackage{graphicx}
\usepackage{soul}
\journalname{}
\begin{document}

\title{Creating a level playing field for all symbols in a discretization.
}
\subtitle{Updating SAX for lowly correlated time-series.}


\author{Matthew Butler         \and
        Dimitar Kazakov
}


\institute{M. Butler \at
            Department of Computer Science,
			The University of York,\\
			Deramore Lane,
			York,
			YO10 5GH,
			United Kingdom \\
              Tel.: +44(0)1904-325500\\
              Fax:  +44(0)1904-325599 \\
              \email{mbutler@cs.york.ac.uk}           
           \and
           D. Kazakov \at
               Department of Computer Science,
			The University of York,\\
			Deramore Lane,
			York,
			YO10 5GH,
			United Kingdom \\
              Tel.: +44(0)1904-325500\\
              Fax:  +44(0)1904-325599 \\
              \email{kazakov@cs.york.ac.uk}  
}

\date{Received: October 2012 / Accepted: date}

\maketitle

\begin{abstract}
In time series analysis research there is a strong interest in discrete representations of real valued data streams. The discretization process offers several desirable properties such as numerosity/dimensionality reduction, the removal of excess noise and the access to numerous algorithms that typically require discrete data. One approach that emerged over a decade ago and is still (along with its successors) considered state-of-the-art is the Symbolic Aggregate Approximation (SAX) algorithm proposed in Lin et al.~\cite{SAX_1}~\cite{SAX_2}. This discretization algorithm was the first symbolic approach that mapped a real-valued time series to a symbolic representation that was guaranteed to lower-bound Euclidean distance. The interest of this paper concerns the SAX assumption of data being highly Gaussian and the use of the standard normal curve to choose partitions to discretize the data. Though not necessarily, but generally, and certainly in its canonical form, the SAX approach chooses partitions on the standard normal curve that would produce an equal probability for each symbol in a finite alphabet to occur. This procedure is generally valid as a time series is normalized to have a $\mu$ = 0 and a $\sigma$ = 1 before the rest of the SAX algorithm is applied. However there exists a caveat to this assumption of equi-probability due to the intermediate step of Piecewise Aggregate Approximation (PAA). What we will show in this paper is that when PAA is applied the distribution of the data is indeed altered, resulting in a shrinking standard deviation that is proportional to the number of points used to create a segment of the PAA representation and the degree of auto-correlation within the series. Data that exhibits statistically significant auto-correlation is less affected by this shrinking distribution. As the standard deviation of the data contracts, the mean remains the same, however the distribution is no longer standard normal and therefore the partitions based on the standard normal curve are no longer valid for the assumption of equal probability.

\end{abstract}

\section{Introduction}
\label{intro}
In time series analysis research there is a strong interest in discrete representations of real valued data streams. The discretization process offers several desirable properties such as numerosity/dimensionality reduction, the removal of excess noise and the access to numerous algorithms that typically require discrete data. As a result there is wealth of literature describing a variety of techniques to facilitate this transformation. These methods can be as simple as equal-width/equal-frequency binning or more sophisticated approaches that are based on clustering~\cite{clustApp} and information theory~\cite{ITA}. One approach that emerged over a decade ago and is still (along with its successors) considered state-of-the-art is the Symbolic Aggregate Approximation (SAX) algorithm proposed in Lin et al.~\cite{SAX_1}~\cite{SAX_2}. This discretization algorithm was the first symbolic approach that mapped a real-valued time series to a symbolic representation that was guaranteed to lower-bound Euclidean distance. This discovery lead to an explosion of application areas for the SAX algorithm, such as, telemedicine~\cite{teleMed}, robotics~\cite{robot}, computational biology~\cite{cb}, environmental science~\cite{es}, network traffic~\cite{nt} and pattern matching in general~\cite{pm1}~\cite{pm2}. The interest in SAX is also motivated by its easy implementation and intuitive approach to discretization based on the assumption of Gaussian distributions. An overview of SAX will follow but at this time we highlight that the interest of this paper concerns the SAX assumption of data being highly Gaussian and the use of the standard normal curve to choose partitions to discretize the data. Though not necessarily but generally and certainly in its canonical form the SAX approach chooses partitions on the standard normal curve that would produce an equal probability for each symbol in a finite alphabet to occur. An equal probability of occurrence for each symbol is considered desirable~\cite{ep1}~\cite{ep2} and supports choosing the partitions to segment the area under the curve into equal regions. This procedure is generally a valid approach as a time series is normalized to have a mean of zero and a standard deviation of one before the SAX algorithm is applied. However there exists a caveat to this assumption of equi-probability, which we will explain in more detail in the following sections, due to the intermediate step of Piecewise Aggregate Approximation (PAA). The PAA step is used for dimensionality reduction and, in brief, it converts a time series into a sequence of means. What we will show in this paper is that when PAA is applied the distribution of the data is altered, resulting in a shrinking standard deviation that is proportional to the number of points used to create a segment of the PAA representation and the degree of auto-correlation within the series. Data that exhibits statistically significant auto-correlation is less affected by this shrinking distribution. As the standard deviation of the data contracts the mean obviously remains the same, since it was zero, however the distribution is no longer standard normal and therefore the partitions based on the standard normal curve are no longer valid for the assumption of equal probability. The rest of this paper is organized as follows: section 2 provides an overview of SAX, section 3 discusses the effects of PAA, section 4 covers the effects of PAA on the symbolic distributions and section 5 demonstrates the relationship between auto-correlation and the effects of PAA.

\section{SAX Overview}
\label{sec:1}

The SAX algorithm maps a real-valued time series of length $\mathit{n}$ to a symbolic representation of length $\mathit{m}$ where $\mathit{m}$ $<$ $\mathit{n}$ and often $\mathit{m}$ $<$$<$ $\mathit{n}$. In this context, $\mathit{m}$ represents the number of segments the time series is divided up into and the ratio of $\mathit{n}$/$\mathit{m}$ could be regarded as the compression rate. The three main steps of the algorithm are as follows:
\begin{enumerate}
\item Normalize the time series to have $\mu$ = 0 and $\sigma$ = 1,
\item convert the time series to PAA, and finally,
\item substitute the PAA segments for symbols corresponding to regions under the standardized normal curve.
\end{enumerate}

The PAA step requires that a time series $\mathit{C}$ of length $\mathit{n}$ is replaced by a vector $\bar{C}$ of length $\mathit{m}$ where each element $\mathit{i}$ of $\bar{C}$ is calculated using the following:
\begin{equation} \label{PAAeq}
\bar{c}_i = \frac{m}{n} \sum^{\frac{n}{m}i}_{j=\frac{n}{m}(i-1)+1} \!\!\!\!\!\!\!\!\!\! c_j
\end{equation}
Equation~\ref{PAAeq} states that a time-series is divided into $\mathit{m}$ equal size segments, where each segment is then represented by its mean. This representation can also be considered as an attempt to approximate the original time-series with a linear combination of box basis functions~\cite{SAX_1}. Figure~\ref{saxGraph1} illustrates the transformation from a real-valued time-series to it's symbolic representation. From the graph we can observe that a series of original length 150 is mapped into a symbolic sequence of 15 letters which form a ``word''~\cite{SAX_1} from an alphabet of cardinality 3. The distance measure, defined on the symbolic sequence, lower bounds the Euclidean Distance of its real-valued counterpart. For example, for two SAX representations $\mathit{P}$ and $\mathit{Q}$ defined over the same alphabet the distance between those sequences is calculated as follows:

\begin{figure} 
\begin{center}
\includegraphics[width=1\textwidth]{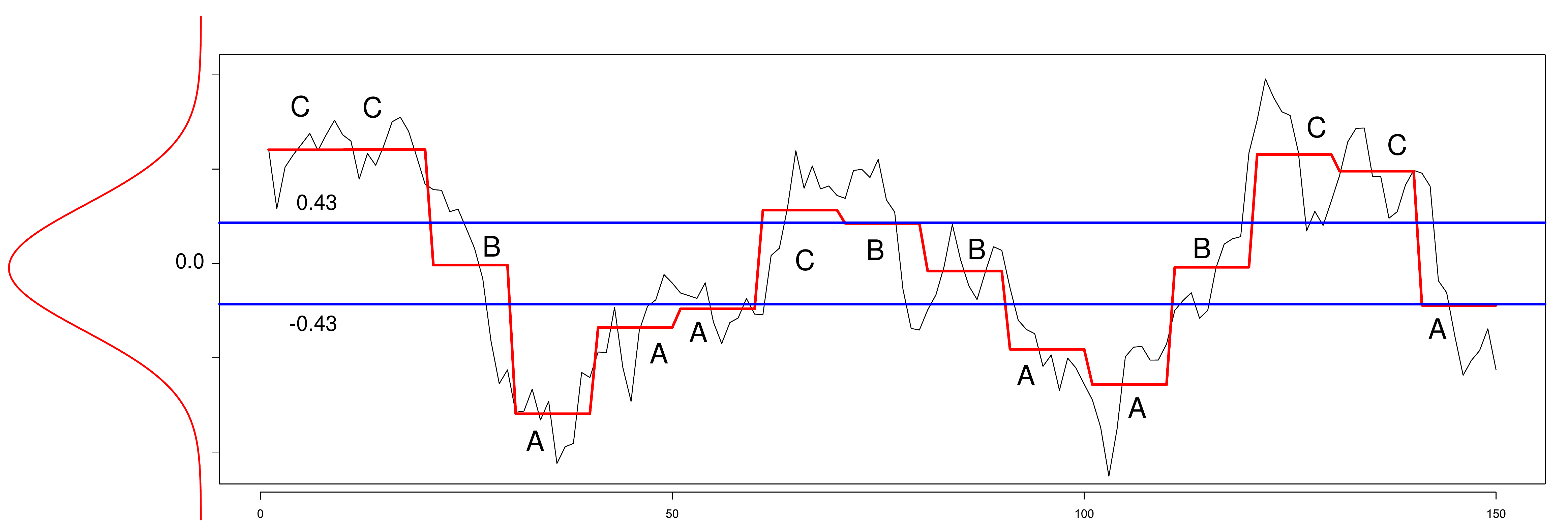}
\caption{A time series of 150 data points converted to a SAX representation of length 15, using an alphabet size of 3. The lines at $\pm$0.43 represent the partitions of the standardized normal curve which yield 3 equally probable regions. The original time series is transformed into the SAX ``word'' CCBAAACBBAABCCA.\label{saxGraph1}}
\end{center}
\end{figure}  

\begin{equation}\label{SAXdistEq}
MINDIST(\hat{Q},\hat{C})\equiv \sqrt{\frac{n}{m}}\sqrt{\sum\nolimits_{i=1}^{m}(dist(\hat{q}_i,\hat{c}_i))^2}
\end{equation}
where the sub function $\mathit{dist()}$ is determined by a lookup table as shown in table~\ref{SAXdistance} for an alphabet of cardinality 4.

\begin{table}
\caption{A lookup table used by the $\mathit{MINDIST}$ function for an alphabet of cardinality 4.}\label{SAXdistance}
\centering\begin{tabular}{c|c|c|c|c|}
\multicolumn{1}{c}{}  & \multicolumn{1}{c}{a} & \multicolumn{1}{c}{b} & \multicolumn{1}{c}{c} & \multicolumn{1}{c}{d} \\ \cline{2-5}
\ a & 0 & 0 & 0.67 & 1.34 \\\cline{2-5}
\ b & 0 & 0 & 0 & 0.67\\\cline{2-5}
\ c & 0.67 & 0 & 0 & 0\\\cline{2-5}
\ d & 1.34 & 0.67 & 0 & 0\\\cline{2-5}
\end{tabular}
\end{table}

\section{Effects of PAA}
\label{sec:2}
As mentioned in the introduction, this paper focuses on the effects of the dimensionality reduction step on the distribution of the data. We are asserting that when PAA is applied to a standard normalized data set that the resulting PAA representation will have a smaller standard deviation. This reduction can be trivial and therefore not affect the assumption of equal probability of each symbol in a finite alphabet. However, depending on the size of the PAA segments and the characteristics of the data this effect can have a significant impact. Trivially, we can highlight this effect by stating that the minimum and maximum of the series will be distorted closer to the mean of the distribution, in all but one special case. This special case occurs when the max and min are surrounded by equal valued points that outnumber or equal the number of points in a PAA segment. To further illustrate this effect we provide examples from simulated and real-world time series. 

\subsection{Simulated Time Series}

\begin{figure} 
\begin{center}
\includegraphics[width=0.75\textwidth]{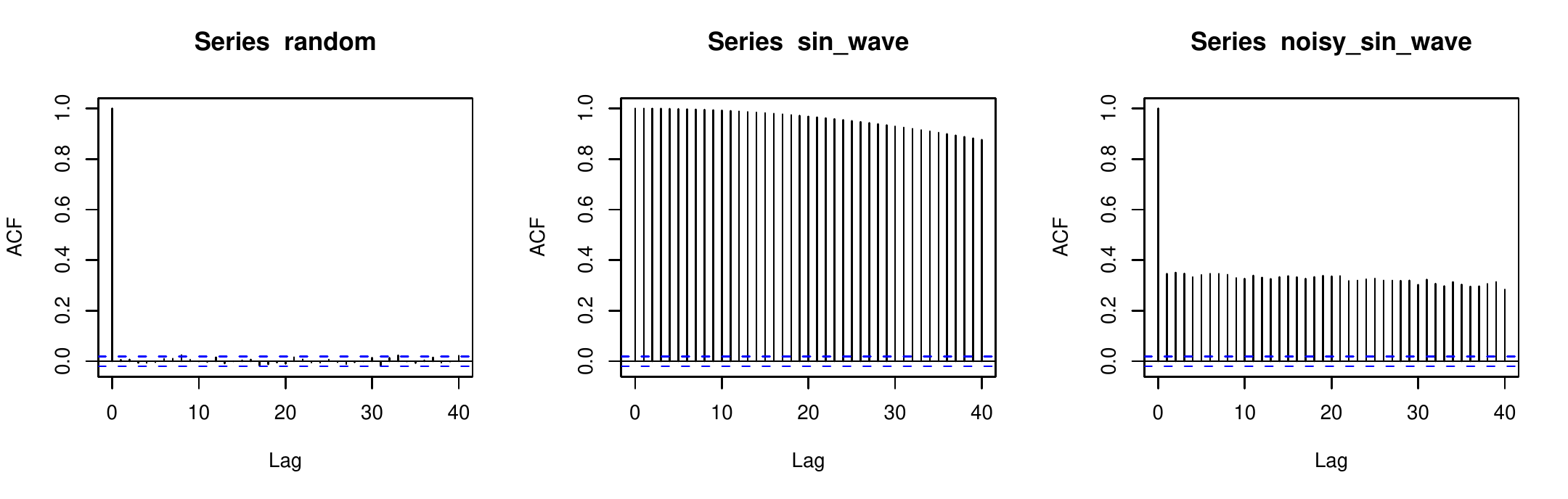}
\caption{Autocorrelation Function (ACF) plots of the three simulated series.\label{ACFplots}}
\end{center}
\end{figure}

This study utilizes three simulated time series which represent the two extreme cases of having highly correlated data (sinusoidal wave form) and completely uncorrelated (white noise) and a mixture of both (sinusoidal wave with added white noise). Figure~\ref{ACFplots} displays the autocorrelation functions for the three series. From these plots we observe that the random data has no significant autocorrelation at any lag, the sinusoidal wave has perfect autocorrelation at lag 1 (which tappers off slowly) and the sinusoidal wave with noise is in between. Additionally, table~\ref{SD4PAA} displays the standard deviations of the simulated time series after the PAA step has been applied for various PAA parameter settings. 

\begin{figure} 
\begin{center}
\includegraphics[width=0.75\textwidth]{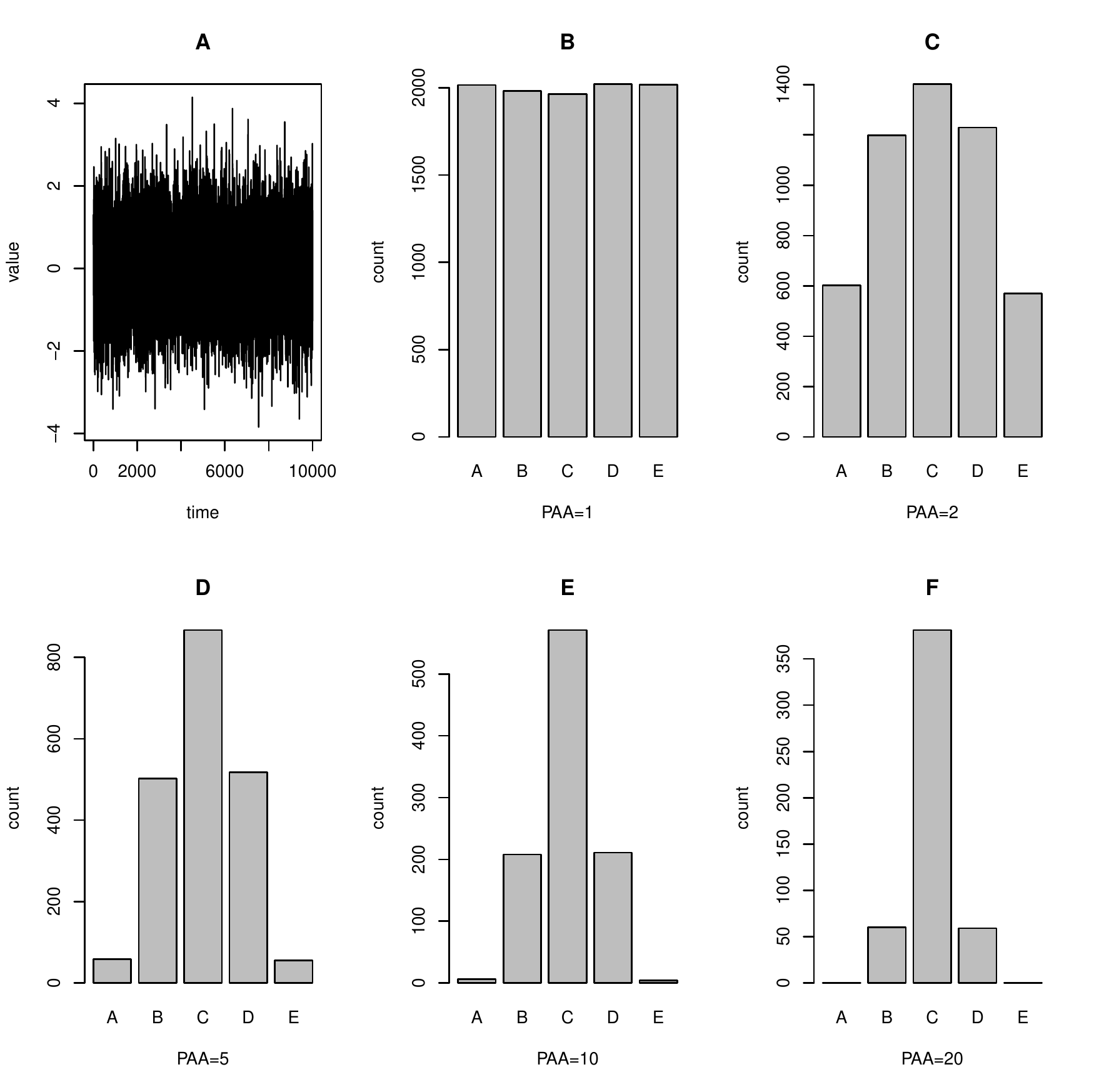}
\caption{Plots related to a standard normal distribution. ``A'' is a time series plot of the data. ``B''-``F'' are box plots of the distributions of letters from a symbolic sequence derived from SAX for increasing number of points used to create the PAA segments. The desired distribution amongst the letters is uniform and is only achieved when the PAA step is skipped (Graph ``B'').\label{randDataBoxPlots}}
\end{center}
\end{figure}

Firstly we have the case were the time series is composed of random data drawn from a standard normal distribution that exhibits no statistically significant autocorrelation. Depicted in figure~\ref{randDataBoxPlots} is a plot of the series and box plots showing the distributions of the letters after the SAX algorithm has been applied. The box plots represent increasing values for the number of points used to construct the PAA segments. Clearly as the number of points increases within a PAA segment the distribution of letters that SAX is mapping to contracts closer to the mean, thus losing the desirable outcome of an equal distribution. The bar chart labelled ``B'' shows the distribution of letters when no dimensionality reduction is applied and is producing a uniform distribution. If we examine the results reported in table~\ref{SD4PAA} we observe that the standard deviation of the distribution is shrinking as the PAA segments become larger.

\begin{figure} 
\begin{center}
\includegraphics[width=0.75\textwidth]{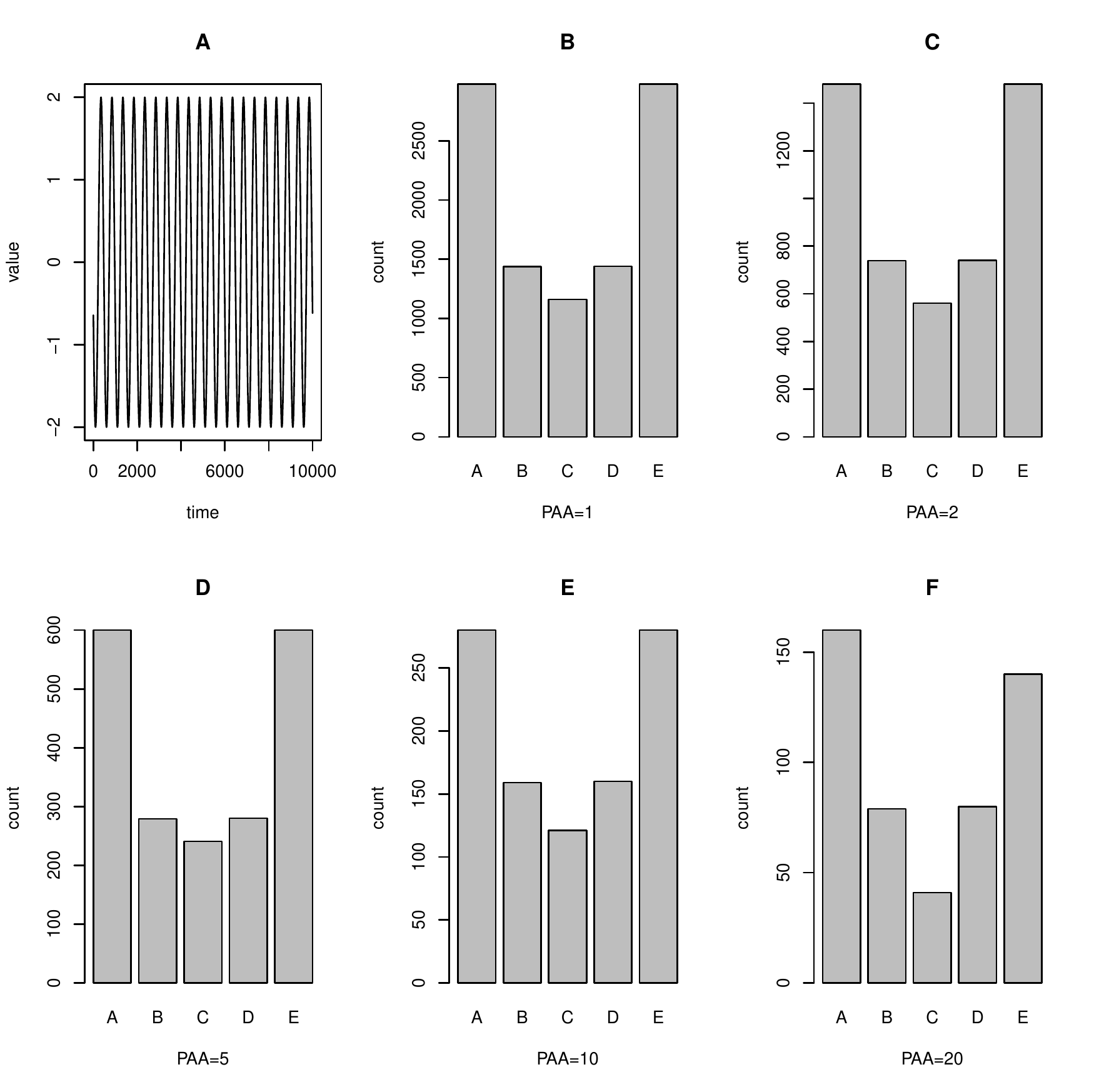}
\caption{Plots related to a sinusoidal wave distribution. ``A'' is a time series plot of the sin wave data. `B''-``F'' are box plots of the distributions of letters from a symbolic sequence derived from SAX for increasing number of points used to create the PAA segments. The desired distribution amongst the letters is uniform and is never achieved as a sin wave is not Gaussian. However, the distribution is never significantly affected by the PAA step.\label{sswNNDataBoxPlots}}
\end{center}
\end{figure}

Secondly we have the case of highly correlated data with the sinusoidal wave form. This data, depicted in figure~\ref{sswNNDataBoxPlots}, was generated using equation~\ref{SSwave}:
\begin{equation} \label{SSwave}
A*cos(2\pi\omega t + \phi)
\end{equation}
where $\mathit{A}$ is the amplitude, $\omega$ is the frequency of oscillation, and $\phi$ is a phase shift, where $\phi$ = $\mathit{B}$ * $\pi$ and $\mathit{B}$ is a constant. For this simulation the values for  $\mathit{A}$, $\omega$, $\mathit{B}$ were 2, 0.002, and 0.6 respectively. From figure~\ref{sswNNDataBoxPlots} we can see that a sin wave is not Gaussian and therefore does not produce a uniform distribution of symbols. However, the data is highly correlated and the PAA step has no effect on the distribution of the data. This result is reflected in the corresponding results in table~\ref{SD4PAA} where the standard deviation is only trivially affected. 

\begin{figure} 
\begin{center}
\includegraphics[width=0.750\textwidth]{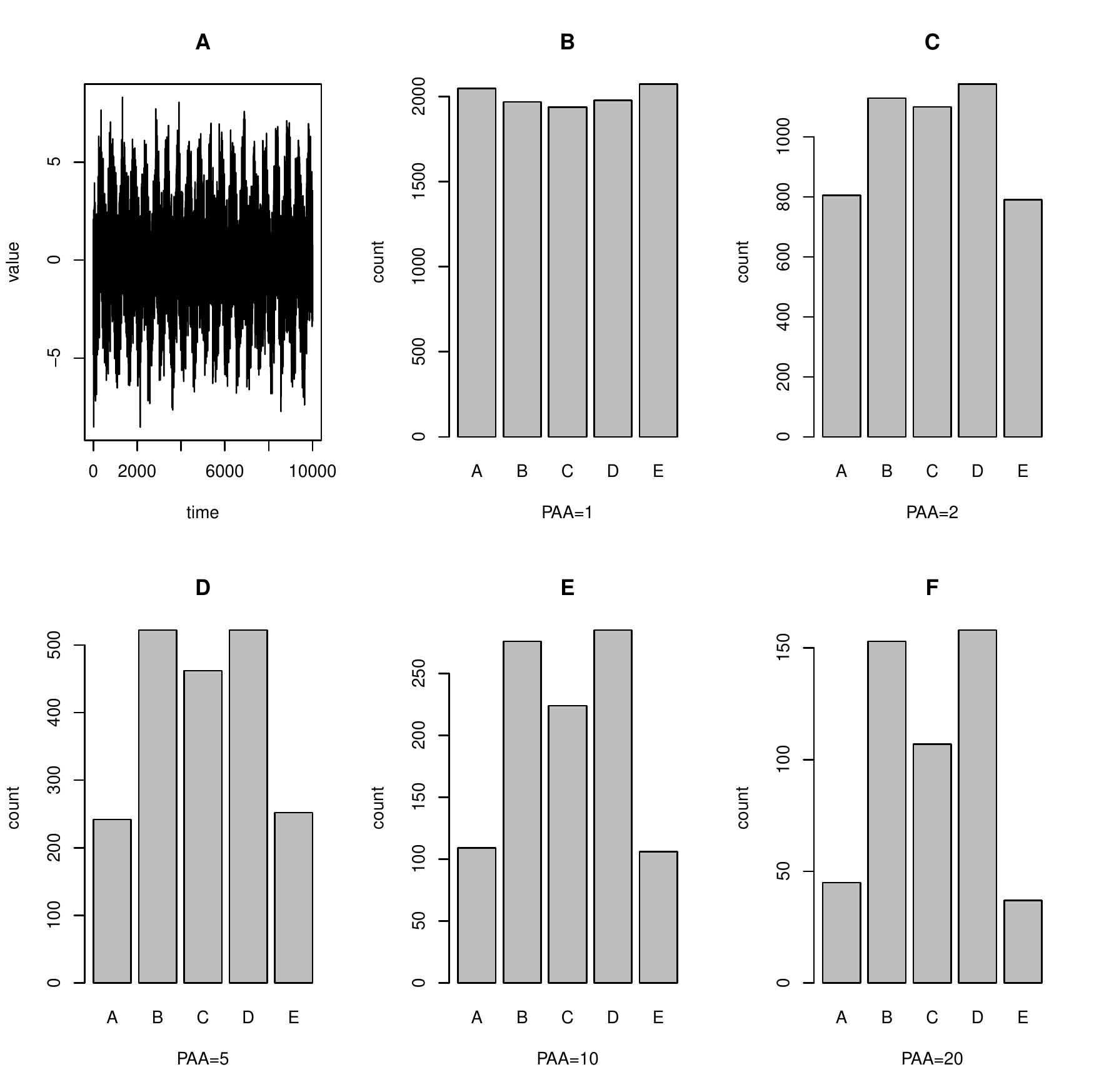}
\caption{Plots related to a sinusoidal wave with added noise distribution. ``A'' is a time series plot of the sin wave data. `B''-``F'' are box plots of the distributions of letters from a symbolic sequence derived from SAX for increasing number of points used to create the PAA segments. The added noise lowers the level of autocorrelation within the sin wave and therefore as PAA is applied the distribution of the data changes.\label{sswWNDataBoxPlots}}
\end{center}
\end{figure}

The final example demonstrates that as the data exhibits lower degrees of autocorrelation the PAA step has a larger impact on the post PAA data distribution. In figure~\ref{sswWNDataBoxPlots} we have the plots of the same sinusoidal wave as before but with added Gaussian white noise.

As we can see from the box plots in figure~\ref{sswWNDataBoxPlots} the data is again not Gaussian but this time the distribution is being altered as the PAA step is applied with more and more data points per segment. The effect is clearly observed in the standard deviations of the ex-post distributions (table~\ref{SD4PAA}) where the standard deviation shrinks from approximately 1 to 0.6098 for PAA segments equal to 20 data points.

\begin{table}
\caption{The standard deviations for the data distributions ex-post the PAA step of SAX.}\label{SD4PAA}
\centering\begin{tabular}{c c c c c c}
\ & 1 & 2 & 5 & 10 & 20 \\\hline
\ random & 0.9999	& 0.7139	& 0.4448	& 0.3167	& 0.2258\\
\ sin wave & 0.9999 & 0.9999 & 0.9998 & 0.9993 & 0.9973\\
\ sin wave with noise &  0.9999	& 0.8162 & 0.6909 &	0.6394 & 0.6098\\
\hline\hline
\end{tabular}
\end{table}

\subsection{Real World Time Series}

\begin{figure} 
\includegraphics[width=1.0\textwidth]{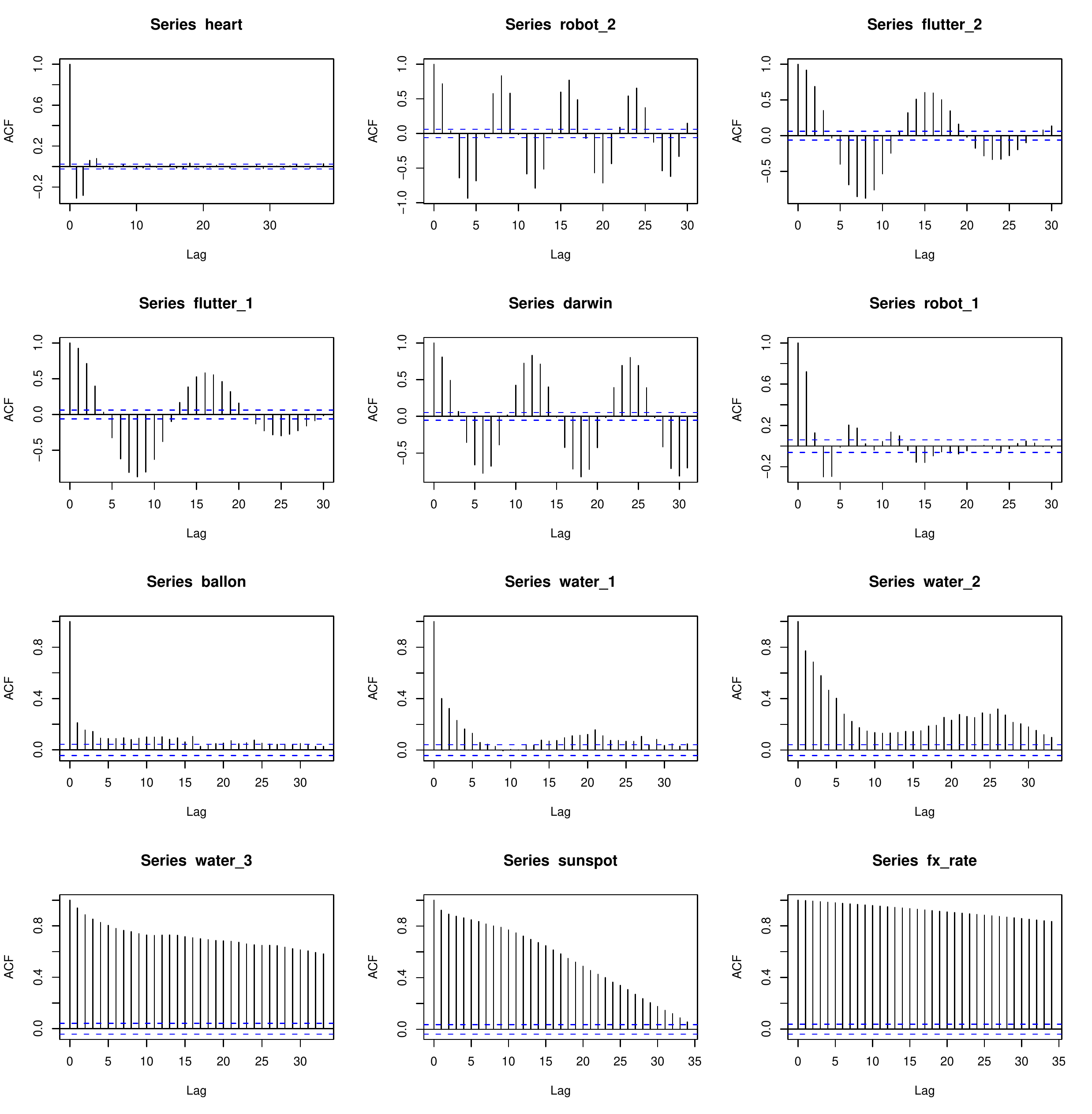}
\caption{Plots of the Autocorrelation functions for the 12 real world time series considered in the study. The ACF plots are arranged based on the impact the PAA step had on the ex-post distribution, where the data series that was most affected is in the top left hand corner and the data set least effected is in the bottom right hand corner.\label{ACF_RW}}
\end{figure}

To demonstrate this effect in real world time series we have chosen 12 data series from 8 sources available from the UCI machine learning repository~\cite{UCI} and downloaded from Dr. Eamonn Keogh's iSAX webpage~\cite{iSAX}. We have chosen 11 series which are negatively impacted by the PAA step and 1 which is not. The 12 data series are summarized in table~\ref{RWdataDes} which reports the results from the Jarque-Bera test for normality and figure~\ref{ACF_RW} displays the ACF plots. Based on the Jarque-Bera test results only one of the time series is normal (robot\_2). The ACF plots are arranged based on the impact the PAA step had on the ex-post distribution, where the data series that was most affected is in the top left hand corner and the data set least affected is in the bottom right hand corner. From these plots we can see that the series with ACFs which are positive and taper off slowly (similar to the sinusoidal wave) were the least affected by the PAA step.

\begin{table}
\caption{The real world time series used to demonstrate the negative effects of the PAA step on the ex-post distributions. JB test stands for the Jarque-Bera test for normality and reports the p-value obtained.}\label{RWdataDes}
\centering\begin{tabular}{l l c c}
\ Data file (name) & Description & length & JB test \\ \hline
\ darwin.dat (darwin) & Monthly values - Darwin SLP series & 1400 & $<$0.001\\
\ flutter.dat (flutter\_1) & Wing flutter data (input) & 1024 & $<$0.001\\
\ flutter.dat (flutter\_2) & Wing flutter data (output) & 1024 & $<$0.001\\ 
\ robot\_arm.dat (robot\_1) & Data from a robot arm (input) & 1024 & $<$0.001\\ 
\ robot\_arm.dat (robot\_2) & Data from a robot arm (output) & 1024 & 0.5042\\ 
\ sunspot.dat (sunspot) & Monthly data - 01/1749 to 07/1990 & 2899 & $<$0.001\\ 
\ EEG\_heart\_rate.dat (heart) & Heart Rate after Epileptic seizure & 7200 & $<$0.001\\
\ water.dat (water\_1) & Rainfall riverflow data (aprecip) & 2191 &  $<$0.001\\
\ water.dat (water\_2) & Rainfall riverflow data (discharg) & 2191 &  $<$0.001\\
\ water.dat (water\_3) & Rainfall riverflow data (Log Flow Rate) & 2191 & $<$0.001\\
\ spot\_exrates.dat (fx\_rate) & Spot prices for GBP in USD & 2567 & $<$0.001\\
\ balloon.dat (balloon) & Balloon collected radiation data & 2001 & $<$0.001\\
\hline
\hline
\end{tabular}
\end{table}

The real world time series are standardized and then converted to a PAA representation. Table~\ref{RW_SDs} displays the ex-post standard deviations of the PAA representations of the standardized time series. As we can see for all time series the standard deviations of the PAA representations are shrinking as the size of the PAA segments increases. However, the spot FX rate exhibits only minor changes in its distribution, thus not impacting the assumption of equi-probability of each symbol. This is similar to the results obtained with the sinusoidal wave without added noise. It is also worth noting that the ACF of the FX rate is similar to the sinusoidal wave as well.  The reduction in the standard deviation in the other 11 series is much more dramatic and therefore more likely to have a negative impact on the ex-post symbolic distribution. This is the same property observed in the simulated sinusoidal wave with noise and the random data drawn from a standard normalized distribution. The ACFs of these 11 series vary in their behaviour but the most affected series are those which have no statistically significant positive autocorrelation at any lag or exhibit oscillating positive and negative autocorrelation.

\begin{table}
\caption{The standard deviations for the data distributions ex-post the PAA step of SAX.}\label{RW_SDs}
\centering\begin{tabular}{l c c c c}
\ & \multicolumn{4}{c}{Size of PAA Segment}\\
\ Name & 2 & 5 & 10 & 20 \\\hline
\ heart & 0.5975	& 0.2246	& 0.1216	& 0.0659\\
\ robot\_2 &0.9268	& 0.5178	& 0.1966	& 0.1404 \\
\ flutter\_2 & 0.9794	& 0.8461	& 0.3982 & 0.1479\\
\ flutter\_1 & 0.9810	& 0.8577	& 0.5665	& 0.1503\\
\ darwin & 0.9496 &	0.7406 & 0.3295 & 0.2607\\
\ robot\_1 & 0.9264	& 0.6449    & 0.4764	& 0.3690\\
\ balloon & 0.7766	& 0.5944	& 0.4690	& 0.3934\\
\ water\_1 & 0.8436	& 0.6683	& 0.5911	& 0.4337\\
\ water\_2 & 0.9461	& 0.8745	& 0.7970	& 0.6157\\
\ water\_3 & 0.9832	& 0.9570	& 0.9283	& 0.8914\\
\ sunspot & 0.9799	& 0.9572	& 0.9341	& 0.8928\\
\ fx\_rate & 0.9994	& 0.9975	& 0.9951	& 0.9875\\
\hline\hline
\end{tabular}
\end{table}

\section{Effect on Symbolic Distributions}

With the property of shrinking distributions now documented in the post PAA representations, the question remains as to what effect this has on the distribution of symbols from a finite alphabet. To demonstrate this effect we assume that a discretization becomes more desirable as its distribution of symbols approaches uniformity. Additionally we introduce a potential fix for this effect, where the PAA data can simply be re-normalized before transformation into a symbolic sequence. Thus the effect can be measured using a chi-squared ($\chi^2$) goodness-of-fit test between the SAX representation and the target uniform distribution. This test will be performed using the canonical form of SAX as well the aforementioned augmented version that renormalizes the PAA representation to have $\mu$=0 and $\sigma$=1. As previously shown (table~\ref{RWdataDes}), only one series can be considered Gaussian ($robot\_2$) and therefore is the only one where we would expect a uniform distribution from the symbolic sequence. For the other 11 series, we would expect that the majority of them will reject the null of the $\chi^2$ test but that the absolute deviations from the uniform distributions should be smaller for a normalized distribution with a $\sigma$=1. Therefore if the shrinking $\sigma$ of the PAA representations is negatively impacting the mapping to a symbolic sequence we should observe smaller absolute deviations and more acceptances of the null of the $\chi^2$ test with the re-normalized PAA representations. The results are reported for the same PAA segmentations as displayed in table~\ref{RW_SDs} and for alphabet cardinalities of 5 and 10. In tables~\ref{chiSqTest1} and~\ref{chiSqTest2} we report the absolute deviations from the uniform distribution for alphabet cardinalities of 5 and 10 respectively; any results which did not reject the null of the $\chi^2$ test (at the 5\% significance level) are marked with an asterisk (*). The null of a goodness-of-fit test is that the observed distribution is equal to the expected distribution. The $\chi^2$ test statistic is calculated as follows:
\begin{equation}
X^2 = \sum^n_{i=1}\frac{(O_i - E_i)^2}{E_i}
\end{equation}
where $\mathit{X}^2$ is the Pearson's cumulative test statistic, $\mathit{O_i}$ is the $\mathit{i^{th}}$ observed frequency, $\mathit{E_i}$ is the $\mathit{i^{th}}$ expected or target frequency and $\mathit{n}$ is the cardinality of the finite alphabet. In these experiments the target frequency would be calculated as 1/(cardinality of the alphabet), so for an alphabet cardinality of 5 the target frequency is 0.2.

\begin{table}
\caption{The results from performing a $\chi^2$ goodness-of-fit test on the SAX symbolic distributions with alphabet cardinality of 5. Reported are the absolute deviations from the uniform distribution. A * indicates the null was not rejected at the 5\% significance level. $SAX_n$ indicates the results when the PAA distribution was re-normalized prior to converting the PAA vector to symbols.}\label{chiSqTest1}
\centering\begin{tabular}{l|c c|c c|c c|c c}
\ & \multicolumn{8}{c}{Size of PAA Segment}\\
\ Name & \multicolumn{2}{c}{2} & \multicolumn{2}{c}{5} & \multicolumn{2}{c}{10} & \multicolumn{2}{c}{20} \\\hline
\ & SAX & $SAX_n$ & SAX & $SAX_n$ & SAX & $SAX_n$ & SAX & $SAX_n$\\\hline
\ darwin & 19.43 & 24.00 & 25.71 & 20.00 & 80.00 & 11.43* & 94.29 & 17.14*\\
\ flutter\_1 & 72.11 &	72.11 &	83.53 &	73.73 &	101.18	& 85.49 &	132.55	& 73.73\\
\ flutter\_2 & 78.75&	77.58	&78.63	&77.65	&105.10&	83.53	&148.24	&81.57\\
\ robot\_1 & 6.17*	&8.52*	&33.92	&14.31*	&62.35	&12.16*	&80.00	&21.96*\\
\ robot\_2 & 13.75	&10.63*	&57.45	&4.51*	&120.78	&9.41*	&144.31	&10.20*\\
\ sunspot & 16.07	&16.48	&16.44	&16.03	&15.64	&13.43*	&17.50*	&14.72*\\
\ heart & 48.28	&3.72*	&110.69	&4.86*	&153.06	&8.33*	&158.89	&8.33*\\
\ water\_1 & 138.81	&118.36	&132.60	&105.66	&129.86	&88.77	&125.14&	77.43\\ 
\ water\_2 & 147.58	&146.85	&147.67	&146.30	&147.21	&143.56	&150.83&	101.65\\
\ water\_3 & 31.78	&31.42	&29.59&	25.94	&27.76	&25.02&	31.93	&24.59*\\
\ balloon & 72.60	&70.40	&73.00	&52.00	&81.00	&34.00	&90.00	&42.00\\
\ fx\_rate & 24.97	&25.13	&25.81	&26.20	&24.53	&25.31	&26.88	&26.88\\
\hline\hline
\end{tabular}
\end{table}

\begin{table}
\caption{The results from performing a $\chi^2$ goodness-of-fit test on the SAX symbolic distributions with alphabet cardinality of 10. Reported are the absolute deviations from the uniform distribution. A * indicates the null was not rejected at the 5\% significance level. $SAX_n$ indicates the results when the PAA distribution was re-normalized prior to converting the PAA vector to symbols.}\label{chiSqTest2}
\centering\begin{tabular}{l|c c|c c|c c|c c}
\ & \multicolumn{8}{c}{Size of PAA Segment}\\
\ Name & \multicolumn{2}{c}{2} & \multicolumn{2}{c}{5} & \multicolumn{2}{c}{10} & \multicolumn{2}{c}{20} \\\hline
\ & SAX & $SAX_n$ & SAX & $SAX_n$ & SAX & $SAX_n$ & SAX & $SAX_n$\\\hline
\ darwin & 30.86&	31.14	&32.14	&22.86	&97.14	&21.43*	&111.43	&34.29*\\ 
\ flutter\_1 & 72.11	&73.52	&83.53&	73.73	&101.18	&88.63	&132.55&	73.73\\
\ flutter\_2 & 78.75	&77.58	&78.63	&77.65	&105.10	&83.53	&148.24	&81.57\\
\ robot\_1 & 7.97*	&9.84*	&36.86	&16.67*	&74.90	&16.47*	&92.55	&41.18*\\
\ robot\_2 & 13.75*	&14.84	&58.24	&12.16*	&122.35	&21.96*	&144.31	&21.96*\\
\ sunspot & 41.48	&42.31	&40.24	&42.63	&39.79	&46.71	&43.06	&45.00\\
\ heart & 49.17	&5.78	&114.86	&4.86*	&153.06	&9.44*	&158.89	&11.67*\\
\ water\_1 & 150.23	&138.36	&138.45	&125.66	&129.86	&109.77	&125.14&	77.61\\
\ water\_2 & 151.69	&151.69	&148.95	&148.95	&147.21	&145.30	&150.83	&125.14\\
\ water\_3 & 44.75	&44.66	&42.47	&42.92	&41.46	&36.99	&41.28	&34.13\\ 
\ balloon & 81.80	&74.60	&78.00	&54.50	&94.00	&45.00	&106.00	&46.00\\
\ fx\_rate & 38.16	&38.32	&40.08	&39.69	&41.88	&41.88	&39.06	&39.06\\
\hline\hline
\end{tabular}
\end{table}

To summarize tables~\ref{chiSqTest1} and~\ref{chiSqTest2}, two statistics can be determined. Firstly, there are 96 cases reported across both tables and in 80 of those cases re-normalizing the data produced a distribution as close or closer to uniform. Secondly, the $\chi^2$ test accepted the null that the two distributions (observed and expected) were equal in 29 of the 96 cases when the PAA distribution was re-normalized but only accepted the null in 4 of the 96 cases when it was not. Based on these results we can conclude that transforming the PAA distribution to standard normal before converting to a symbolic representation, facilities a more even distribution of the symbols. However, in the case of the FX rate, re-normalizing lead to poorer results from the $\chi^2$ test in 4 of the 8 cases. In the 2 cases that a better result was achieved, it was only a marginal improvement. From these results we can conclude that, whether or not the effect of the PAA step can be determined a priori, a simple test of the PAA distribution is all that is required to decide if a re-normalization is necessary. Clearly in the case of the FX rate the re-normalization was not advantageous.

If we focus on robot\_2, the only series that was Gaussian, we observe similar results to those obtained with the simulated data from the standard normal distribution. The $\sigma$ steadily shrank and the absolute deviation from the uniform distribution grew as the size of the PAA segments became larger. When we examine the re-normalized PAA representation we see that in 7 of the 8 cases the symbolic sequence produced did conform to a uniform distribution and in the one case that the $\chi^2$ test rejected the null, it was marginal.

\section{Effect of Autocorrelation}

\begin{figure} 
\includegraphics[width=1.0\textwidth]{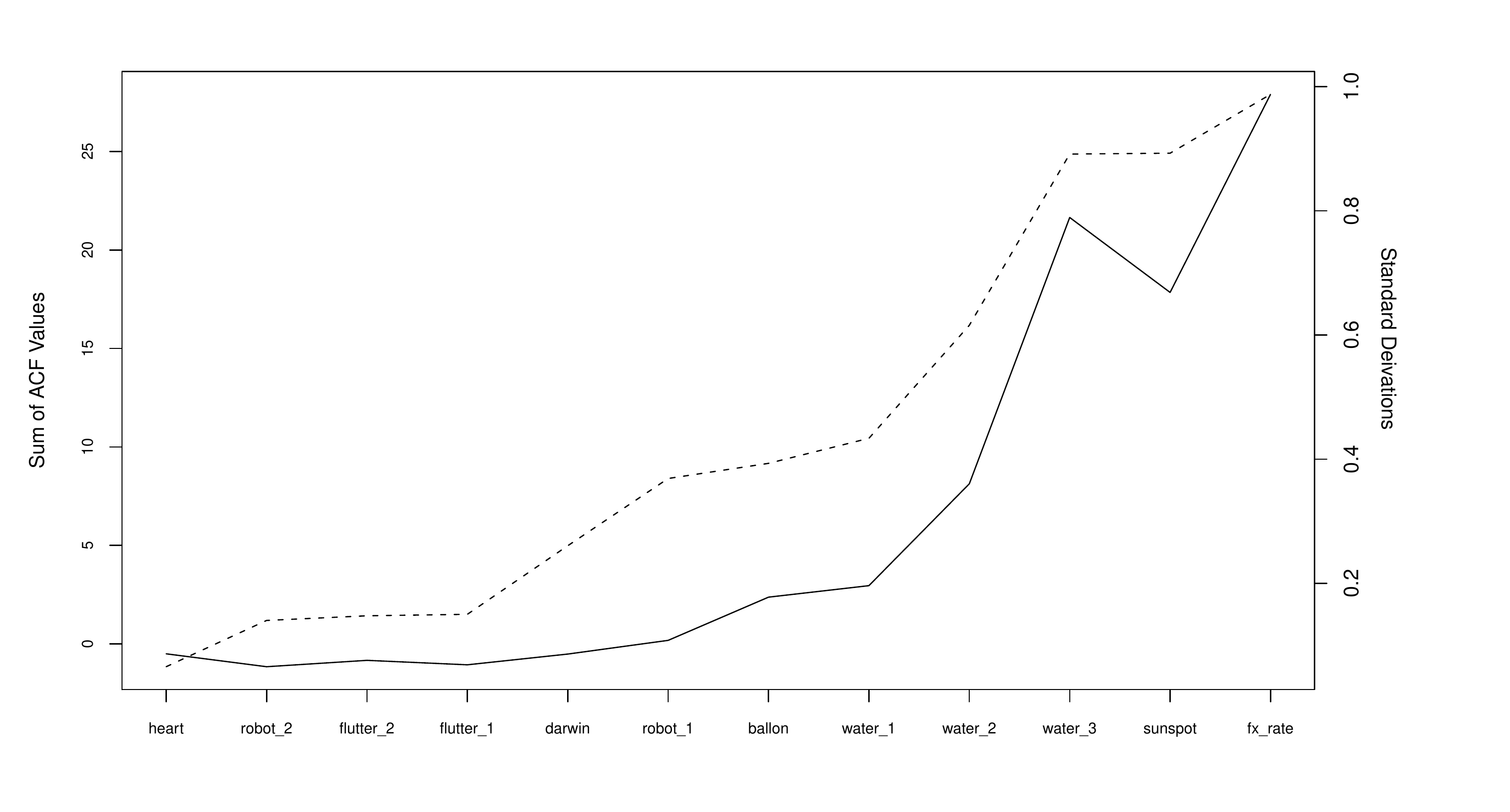}
\caption{Plots of the standard deviations (dashed line) and ACF coefficient sums (solid line) for the 12 data series under analysis. The two plots reveal a strong linear relationship between the variables which is also expressed by the correlation coefficient between them of 0.9527805.\label{ACFsumandSDplots}}
\end{figure}

We are asserting that autocorrelation is a non-trivial contributing factor to the effect of the PAA step on the standard deviation of the PAA representation distribution. Figure~\ref{ACF_RW} displays the plots the ACFs of the time series in order of effect the PAA step had on the standard deviation for the PAA segment size of 20. This figure depicted a succession of ACF plots that became more and more positively autocorrelated as the effect on the standard deviation diminished. To further highlight this point in figure~\ref{ACFsumandSDplots} we provide a plot of the sums of the autocorrelation coefficients up to 30 lags for the 12 series along with a plot of the standard deviations. In figure~\ref{ACFsumandSDplots} the solid line represents the sum of the ACF coefficients and the dashed line represents the standard deviations. The two plots are clearly linearly correlated, where an increase in the ACF sum is related to a decrease in the effect on the post PAA distribution. This relationship is also expressed by the correlation coefficient between the two series, which is 0.9527805. Additionally, if we fit a linear regression model of the form:
\begin{equation}
Y = X\beta + \epsilon
\end{equation}
where $\mathit{Y}$ (the response variable) represents the $\sigma$ and $\mathit{X}$ represents the ACF coefficient sums as the predictor. The resulting adjusted $R^2$ value is 0.8989, thus indicating that approximately 90\% of the variance in the post PAA standard deviations is explained by the autocorrelation. 

\section{Conclusions}

In this paper we highlighted conditions under which one of the core assumptions of the SAX algorithm is invalid. When the PAA step is applied for dimensionality reduction the standard deviation of the resulting distribution is almost certainly less than 1 and dependent on the circumstances can be much less than 1. Those circumstances concern two measurable variables, the number of data points within a given PAA segment and the degree of positive autocorrelation within the series. A time series with statistically significant autocorrelation which tappers of slowly will be less effected by the PAA step and vice-versa. We have shown that the shrinking distribution negatively effects the symbolic representation of the time series with respect to the target uniform distribution. Finally we have proposed a small change to the SAX algorithm which entails testing the standard deviation of the PAA distribution and re-normalizing if it has been reduced below some desirable threshold. Implementing this additional step will allow a very popular algorithm to realize its full potential in the variety of domains where it has already proved very useful.

\bibliographystyle{spmpsci}
\bibliography{newSAXreferences}

\end{document}